# Binar Sort: A Linear Generalized Sorting Algorithm

by

William F. Gilreath

February 2008

(will@williamgilreath.com)

Abstract

Sorting is a common and ubiquitous activity for computers. It is not surprising that there exist a plethora of sorting algorithms. For all the sorting algorithms, it is an accepted performance limit that sorting algorithms are linearithmic or $O(N \lg N)$. The linearithmic lower bound in performance stems from the fact that the sorting algorithms use the ordering property of the data. The sorting algorithm uses comparison by the ordering property to arrange the data elements from an initial permutation into a sorted permutation.

Linear $O(N)$ sorting algorithms exist, but use a priori knowledge of the data to use a specific property of the data and thus have greater performance. In contrast, the linearithmic sorting algorithms are generalized by using a universal property of data-comparison, but have a linearithmic performance lower bound. The trade-off in sorting algorithms is generality for performance by the chosen property used to sort the data elements.

A general-purpose, linear sorting algorithm in the context of the trade-off of performance for generality at first consideration seems implausible. But, there is an implicit assumption that only the ordering property is universal. But, as will be discussed and examined, it is not the only universal property for data elements. The binar sort is a general-purpose sorting algorithm that uses this other universal property to sort linearly.

## Introduction

Sorting is a frequent, ubiquitous computational activity on computers, often in tandem with other algorithms such as search. The interest and ubiquity of sorting creates has led to the development of a diverse number of sorting algorithms. But it begs the question of “What exactly is sorting?”

## Sorting Formalism

Knuth [Knuth 1998] defines sorting as: The records

$$R_1, R_2, \ldots , R_N$$

are supposed to be sorted into nondecreasing order of their keys $K_1, K_2,\ldots,K_N$, essentially by discovering a permutation p(1) p(2)…p(N) such that

$$K_{p(1)} \leq K_{p(2)} \leq \ldots \leq K_{p(N)} .$$

Knuth’s definition of sorting defines sorting in terms of the end result, a specific permutation out of the *n!* possible permutations possible for a collection of records. The definition is more mathematical than algorithmic-what is the result is described, but now the how. The how is the specific sorting algorithm.

## Linearithmic Sorting

A sort algorithm performs sorting, and most sorting algorithms are comparison based sorting algorithms. The comparison sorting algorithms include such algorithms as the merge sort, quick sort, and heap sort. These sorting algorithms use comparison to arrange the elements in the sorted permutation, and are general-purpose in nature.

The comparison sorting algorithms have a well-known theoretical [Johnsonbaugh and Schaefer 2004] performance limit that is the least upper bound for sorting that is linearithmic or $O(N \lg N)$ in complexity. This theoretical lower bound is from the basis of the comparison sorting algorithm-using sorting to arrange the data elements. A decision tree for $N$ elements is of logarithmic height $\lg N$. Thus the time complexity involves the cost that is the cost of using a decision tree to compare elements $\lg N$ and the number of elements $N$. Hence the theoretical least upper bound is the product of the two costs involved, $O(N \lg N)$, which is linearithmic complexity.

## Linear Sorting

Most, but not all sorting algorithms are linearithmic and use comparison as the main operator to arrange the data elements. But there are sorting algorithms that are linear, or $O(N)$ complexity but are not based on utilizing comparison. Such linear algorithms include the radix sort, hash sort, and counting sort.

Linear sorting algorithms use a priori knowledge of the data elements, a specific property is required to sort linearly. The linear sorting algorithms map a data element using the specific property to formulate a function or operation that maps the data.

The radix sort operates on a known radix or base. The hash sort on numbers within a specific range, and the counting sort a known range of elements. But for each linear sorting algorithm requires a priori knowledge about the data set to utilize the specific property.

Thus the linear sorting algorithms are not general-purpose as they are tied to a specific property. Compared (no pun intended) to the linearithmic sorting algorithms, the linear are special-purpose. In sorting algorithms, there is the classic engineering trade-off of generality with performance for linear with linearithmic, respectively.

## Summary of Linear with Linearithmic

Sorting is algorithmic, but the algorithm has a key operation based upon a property of the data elements. Linearithmic sorting algorithms use comparison, or a comparison operation, but that operation is based upon the universal property of ordering. Linear sorting algorithms use a mapping operation based upon a specific property of the data elements.

At first glance, it would seem that the trade-off in generality to performance for sorting algorithms is an absolute rule. But such a rule makes an implicit presumption that there is no other universal property of the data elements. The implicit presumption is that ordering is the only universal property of data elements. But in point of fact ordering is not the only universal property of data elements.

## The Other Universal Property

It should be noted that ordering in the form of a comparison function or operator is not as universal as it seems. Frequently in object-oriented software development, a class must define a comparison method to determine the ordering of instances of the class as objects. From a mathematical perspective, numbers such as real numbers and integer numbers do have ordering, but from an algorithmic perspective the ordering is not always cleared defined or given. This hints that ordering is not the only universal property, and thus the trade-off in sorting algorithms of performance to generality is not a hard, fast rule.

The question is then one of "What is the alternate universal property of data?" The answer is seemingly obvious, but no so—the encoding of the data, which is how a data element is represented in the computer.

The basic data elements of integers, doubles, floats, and characters are all encoded in a binary form on a computer. Computers are binary in operation (which is a given de facto standard of current technology, but is not an absolute) and the representation or encoding is in bits--binary digits. But for computers to exchange and share data the representation must be agreed upon--the encoding.

There are different encoding standards used, such as ASCII code [ANSI 1986], UTF-8 [Yergeau 2003], and Unicode [Unicode 2006] for characters, IEEE-754-1985 [IEEE 1985] for floating point numbers, and two's complement for integers. Encoding is universal because computers require an ubiquitous form of representation to inter-change data in more complex forms. But the encoding is frequently seen as a hardware specific detail, not a software, which is why encoding, is not readily apparent as a universal property. Data automatically exists in the computer, and the details of the data are a hardware, not software concern.

Given that the encoding is universal, "How does this relate to a sorting algorithm?" is the germane question. Simply put, finding another universal property allows for generality, and encoding is not ordering, hence the algorithm is not comparison sorting. Thus the linearithmic lower bound does not necessarily apply. Utilizing the encoding, a sorting algorithm can map the data elements instead of compare them, which is the basis for a general-purpose linear sorting algorithm. A function or operator is needed to use the encoding to arrange the data elements from an initial random permutation into a sorted permutation. Such a sorting algorithm is the binar sort algorithm.

Algorithm Synopsis

The binar sort places data elements from the initial array into two sub-arrays called the lower and upper sub-arrays. (This process is termed partitioning of an array into sub-arrays.) The algorithm extracts the nth bit from a position within a given data element, and then uses that bit (which is either zero or one) to place the element in the lower or upper sub-array.

The key operation in the binar sort is the bit extraction, and then the use of the bit to put the data element in the correct sub-array depending upon the bit value. (for each bit what partition explanation?)

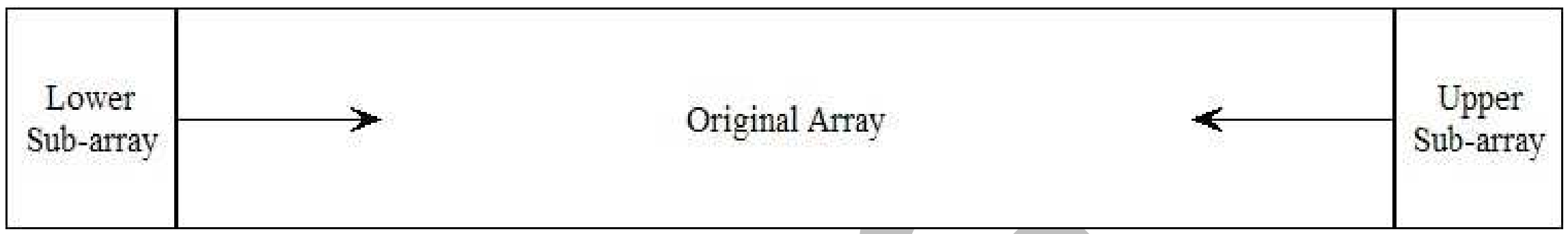

Diagram of Two Sub-Arrays at Extreme Ends of the Original Array

For each bit extracted, one or two sub-arrays are created depending upon the bit values, but in either case the bit extraction continues on the resulting sub-arrays.

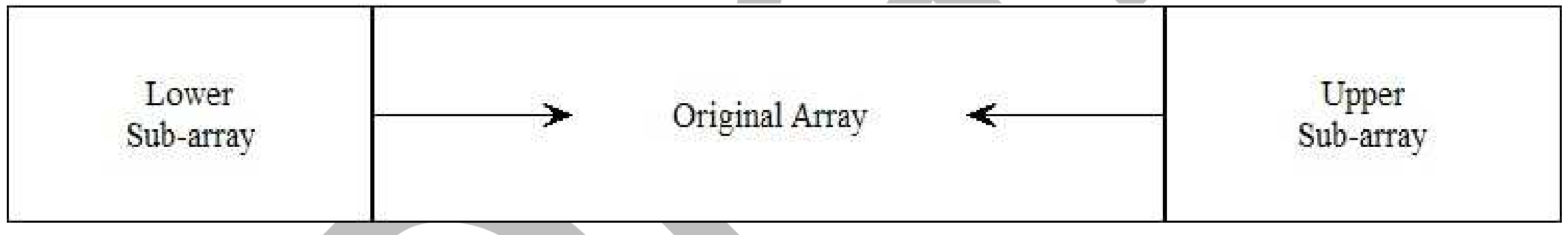

Diagram of Two Sub-Arrays Approaching One Another

The two sub-arrays are created within the original array starting at the extreme endpoints (for an array of $N$ elements, index positions 0 to N-1) and progressing towards each other at the middle of the initial array.

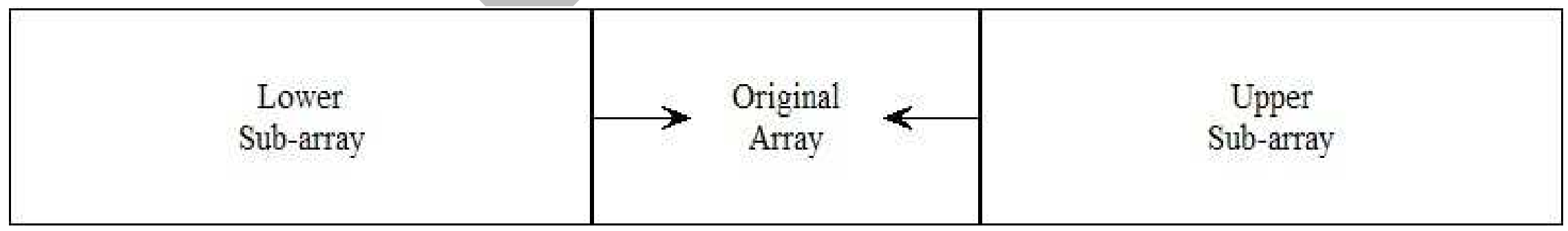

Diagram of Two Sub-Arrays Encompassing Original Array Approaching One Another

When the sub-array boundaries cross or intersect, the process then continues on each created sub-array using the next bit position in the data elements.

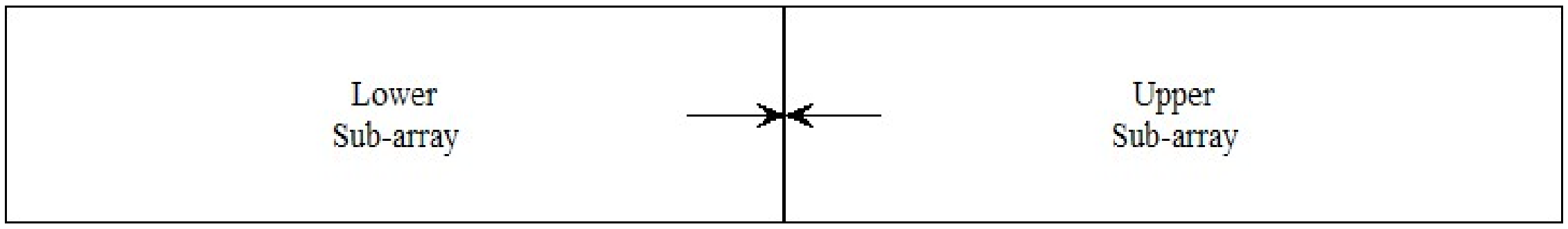

Diagram Illustrating Two Sub-Arrays at Intersection or Crossover

This process continues on each sub-array repeatedly by recursion until either the last bit in the data element is reached, or the size of a sub-array is one element. At the terminating case either the data element is in a final sub-array by itself; it cannot be partitioned into more sub-arrays, or there are no more bits to use to arrange the data elements. The original array is now in a sorted permutation and thus the binar sort algorithm is finished.

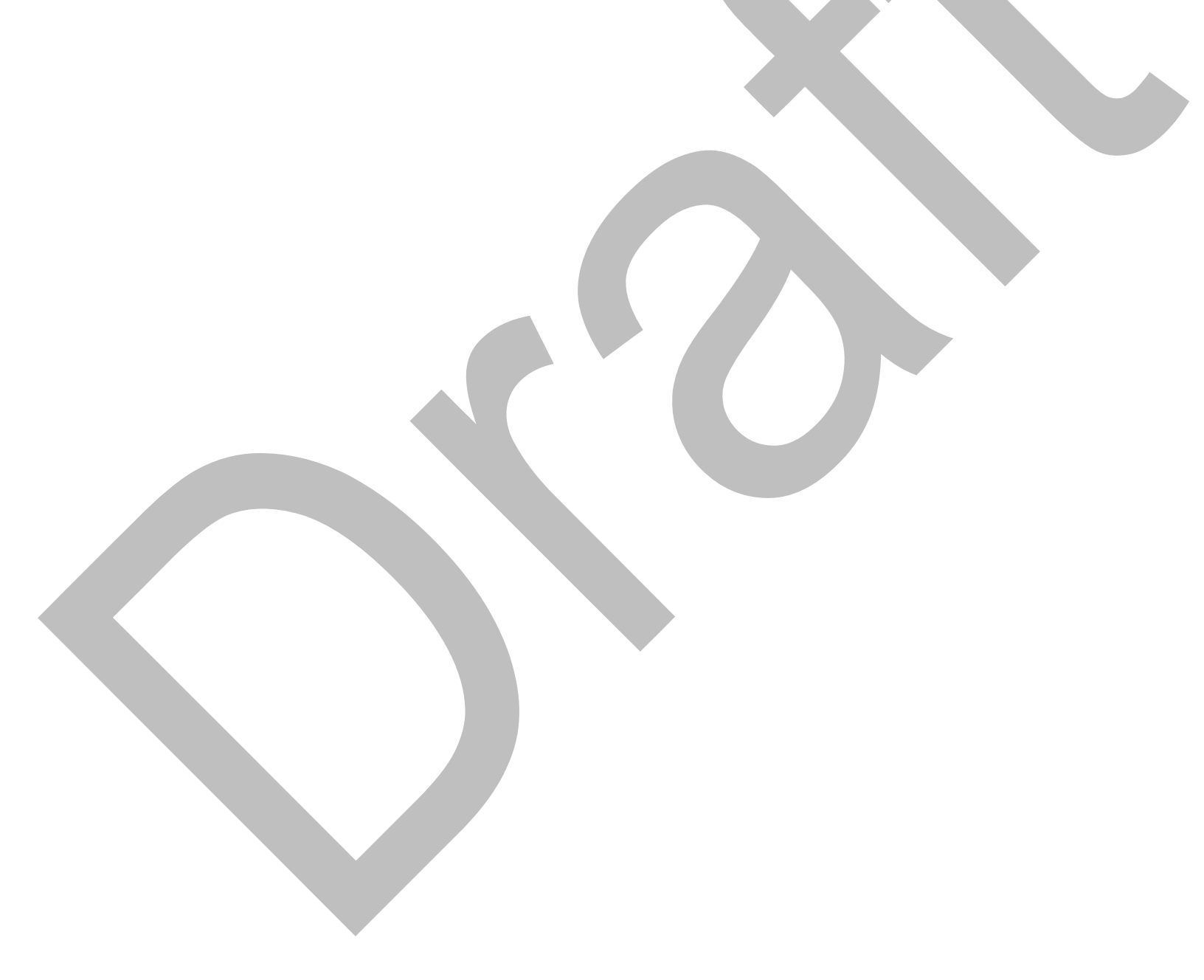

## Operation of Algorithm

The binar sort algorithm operates both iteratively and recursively in place on the original array passed as a parameter initially. The algorithm operates in four discrete steps, which are:

1. Evaluate for recursive base case.
2. Initialize starting array bounds.
3. Partition array into sub-arrays.
4. Determine recursive call on sub-arrays.

### Evaluate for Recursive Base Case

The first step is to evaluate the passed parameters for termination of the algorithm, to determine if a recursive base case has been reached. When one of the two possible base cases of the binar sort is reached, the recursion terminates, and the call returns without any further operation on the passed parameters. The two criteria for the base case of the recursion are:

1. Reach the end of the bits in an element.
2. The size of a sub-array is one element.

Both cases are simple enough. The first case is to reach the end of the number of bits for a data element. In effect, there are no more bits to extract to use for partitioning. The second case the bounds of the array parameter are evaluated to see if there are any elements to partition into a sub-array. For an array of one element there is no point to partition, as the element is in its final position. Thus for either case the operation of the binar sort terminates, or the recursive method call returns.

### Initialize Starting Array Bounds

If the binar sort algorithm does not terminate, then the operation proceeds to the initialization step. The passed array bounds are used to initialize the bounds for the original array. The passed array bounds are retained for use later in the operation of the algorithm. The bounds of the original array are used by variables to track the changing boundaries of the original array during partitioning. After initializing the original array boundaries, the operation then proceeds to partition.

### Partition Array into Sub-arrays

The heart of the binar sort algorithm and the key operation is the partitioning of the original array. The partition step of the algorithm divides the elements of the original array into lower and upper sub-arrays. The partition operation has three distinct steps:

1. Extract the nth bit from the data element in selected position.
2. Using the bit value, place the data element in the correct sub-array.
3. Adjust array boundary so the correct sub-array is extended to encompass element.

It is important to note that the selected position used in the operation is the lower position of the array. The selected position acts as a point of focus for the operation of partitioning the array, the data element at the lower position is the working element, the element that is being placed into a sub-array by partitioning.

Bit Extraction

Before any partitioning is possible on the selected data element, the nth bit must be extracted to determine the sub-array to place the element. The process of bit extracting uses a shift operation to the left by n-bits, and a bit mask to extract the bit as an integer zero or non-zero (not necessarily integer value of 1, but the integer value of the bit mask literal). The bit mask is a literal that depends on the data element word size. The bitwise logical and operation masks all the bits to zero or to the value of the literal bit mask.

For example for a data element of 4-bits, or a nybble, the bit mask is hexadecimal `0x1000`, or an integer value of 8. For a data element the bit mask with a logical bitwise and operation is:

```
0xXXXX && 0x1000 → 0xX000
```

The resulting value for a nybble is either hexadecimal `0x0000` the integer 0, or hexadecimal `0x1000` the integer 8, or a non-zero. For bit extraction, the result of the extracted bit for a data element is essentially a zero or non-zero value.

Placement of Element

The placement of the data element is dependent upon the bit value from the bit extraction. Depending on the bit value, the data element is placed by one of two possibilities. The data element selected is in the lower sub-array. The two possibilities are:

1. The data element is already placed in correct position in the lower sub-array.
2. The data element is in wrong position, exchange with the element in the upper sub-array.

For a bit value of zero, the data element is in the correct position in the lower sub-array, so nothing is done. For a bit value of non-zero, the data element is exchanged or swapped with the data element in the upper sub-array. In both cases, it then follows to adjust the array bounds after the data element is placed.

Adjust Array Bounds

Once the data element is placed in the correct sub-array, the array boundaries are adjusted to encompass the element. For the lower sub-array, the lower array bound is incremented, and for the upper sub-array, the upper array bound is decremented. In the operation of the algorithm, the lower and upper array bounds approach one another as each data element is placed in the correct sub-array.

Partition Repetition

The partitioning process continues iteratively for each data element to be correctly placed in a sub-array. The iterative process continues until the array bounds cross over or overlap. At which points all the elements are partitioned into the sub-arrays. The last step is to determine the recursive call to continue the operation of the algorithm on the sub-arrays.

Determine Recursive Call on Sub-arrays

The last primary step of the operation of the binar sort is to continue the algorithm recursively on the sub-arrays. Depending upon the process of partitioning, it is possible that one or two sub-arrays were created. For one sub-array, no partitioning occurred; effectively the original array is undivided. The condition of not partitioning the original array into two sub-arrays is termed pass-through. With two sub-arrays, the partitioning process successfully divided the data elements of the original array into two sub-arrays.

However, the operation of the algorithm must determine which case is the result of partitioning--pass-through with one sub-array, or two sub-arrays. The original passed array bounds are evaluating using the array bound variables that changed during partitioning. From the evaluation the recursive call is either a single recursive call or two recursive calls. The parameters passed involve the original array bounds, the array variables modified during partitioning, and the original array. For either recursive call the bit position is the next incremental bit position in the data element from the current position $i$ to the next position $i+1$.

## Illustration of the Algorithm

The binar sort algorithm works on different data types (such as character, integer, ordinal, float, double, string) of different data sizes. The word size of the data element is a constant of the algorithm. In demonstration of the algorithm, a 4-bit size word or nybble (a single hexadecimal character or digit) is used for illustration.

Consider an initial array of nybble values `[B 4 0 0 7 A C E]`, which are eight nybbles or hexadecimal characters to sort into an ordered permutation. The bit mask for the bit extraction is hexadecimal `0x1000`, or integer value of 8.

## First Pass at Partitioning

The first pass at partitioning divides the array (in the center) into a lower sub-array (on the left) and an upper sub-array (on the right). Initially before any partitioning the configuration of the array, lower sub-array, and upper sub-array is:

```
[][B 4 0 0 7 A C E][]
```

The selected element is 'B', and the bit extracted is from the most significant position, which is a non-zero. The element is not in the correct position, so is swapped with the element at the boundary of the upper sub-array, the 'E' element. After the exchange, and array boundary adjustment the configuration is:

```
[][E 4 0 0 7 A C][B]
```

The selected element is 'E', and the bit extracted is non-zero. Again, the element is not in the correct position, so is swapped with the 'C' element. The configuration is:

```
[][C 4 0 0 7 A][E B]
```

The selected element is 'C', and the bit extracted is non-zero. The element is not in the correct position, and so is swapped the 'A' element. The configuration is:

```
[][A 4 0 0 7][C E B]
```

The selected element is 'A', and the bit extracted is non-zero. The element is not in the correct position, and so is swapped the '7' element. The configuration is:

```
[][7 4 0 0][A C E B]
```

The selected element is '7', and the bit extracted is zero. The element is in the correct position, and so only the array bounds are adjusted. The configuration is:

```
[7][4 0 0][A C E B]
```

The selected element is '4', and the bit extracted is zero. Again, the element is in the correct position, and so only the array bounds are adjusted. The configuration is:

```
[7 4][0 0][A C E B]
```

The selected element is '0', and the bit extracted is zero. The element is in the correct position, and so only the array bounds are adjusted. The configuration is:

```
[7 4 0][0][A C E B]
```

Again, as the selected element is '0', and the same process with the final configuration of:

```
[7 4 0 0][A C E B]
```

The lower sub-array is [7 4 0 0] and the upper sub-array is [A C E B]. The algorithm continues on each sub-array as an array, further partitioning them on the next bit position.

Partition from Initial Lower Sub-Array

The partitioning of the lower sub-array is on the second bit position. As before the initial configuration is:

```
[][7 4 0 0][]
```

The selected element is '7', and the extracted bit is non-zero. The element is placed with an exchange, and the configuration is:

```
[][0 4 0][7]
```

The selected element is '0', and the extracted bit is zero. The element is in the correct position, and the configuration is:

```
[0][4 0][7]
```

The selected element is '4', and the extracted bit is non-zero. The element is placed with an exchange, and the configuration is:

```
[0][0][4 7]
```

The selected element is '0', and the extracted bit is zero. The element is in the swapped into position (with itself as there is only one element), and the configuration is:

```
[0 0][][4 7]
```

The resulting lower and upper sub-array elements are in the correct positions. However, partitioning would continue, but in the case of pass through resulting in only one sub-array. The remaining third and fourth bit positions are evaluated, and when the last bit position is reached.
Partition from Initial Upper Sub-Array

The partitioning of the lower sub-array is on the second bit position. As before the initial configuration is:

```
[][A C E B][]
```

The selected element is 'A', and the extracted bit is zero. The element is in the correct position, and the configuration is:

```
[A][C E B][]
```

The selected element is 'C', and the extracted bit is non-zero. The element is swapped into the correct position, and the configuration is:

```
[A][B E][C]
```

The selected element is 'B', and the extracted bit is zero. The element is in the correct position, and the configuration is:

```
[A B][E][C]
```

Lastly, the selected element is 'E', and the extracted bit is non-zero. The element is swapped (with itself as there is only one element) and the array bounds adjusted. The final configuration is:

```
[A B][][E C]
```

The lower sub-array for the third and fourth bits the elements are in the correct position, the case of pass through. For the upper sub-array, the elements are positioned on the third bit, but then the recursion terminates, as the sub-array size is one element.

Summary of Illustration

The partitioning process starts from an initial array of hexadecimal characters of 4-bits, and continues until the end of the last bit (pass through) or when the sub-array is only of one element in size. Consider the array configuration into sub-arrays for each bit position, starting from an initial array. The configuration for each pass is:

1. Beginning of Array: `[B 4 0 0 7 A C E]`
2. Partition on Bit 1: `[7 4 0 0]    [A C E B]`
3. Partition on Bit 2: `[0 0] [4 7] [A B]  [E C]`
4. Partition on Bit 3: `[0 0] [4 7] [A B] [C] [E]`
5. Partition on Bit 4: `[0 0] [4 7] [A B] [C] [E]`

Note from partitioning bit 2 to bit 3, the sub-array with `[E C]` is partitioned into two sub-arrays of a single element. Thus for the remaining bit positions, no partitioning occurs as the recursion terminates. However the other sub-arrays are the case of pass through, no data elements are parti-tioned. This continues until the last bit position is reached, and the recursion terminates. Both recursive cases are used in the overall partitioning process.

After the last bit position is evaluated for the three sub-arrays that are partitioned but result in pass through, the overall array is in the final configuration of:

```
[0 0 4 7 A B C E]
```

The configuration is by Knuth's definition [Knuth 1998] a sorted permutation of the array. The partitioning on the sub-arrays operates in place, within boundaries for sub-arrays within the original overall array.

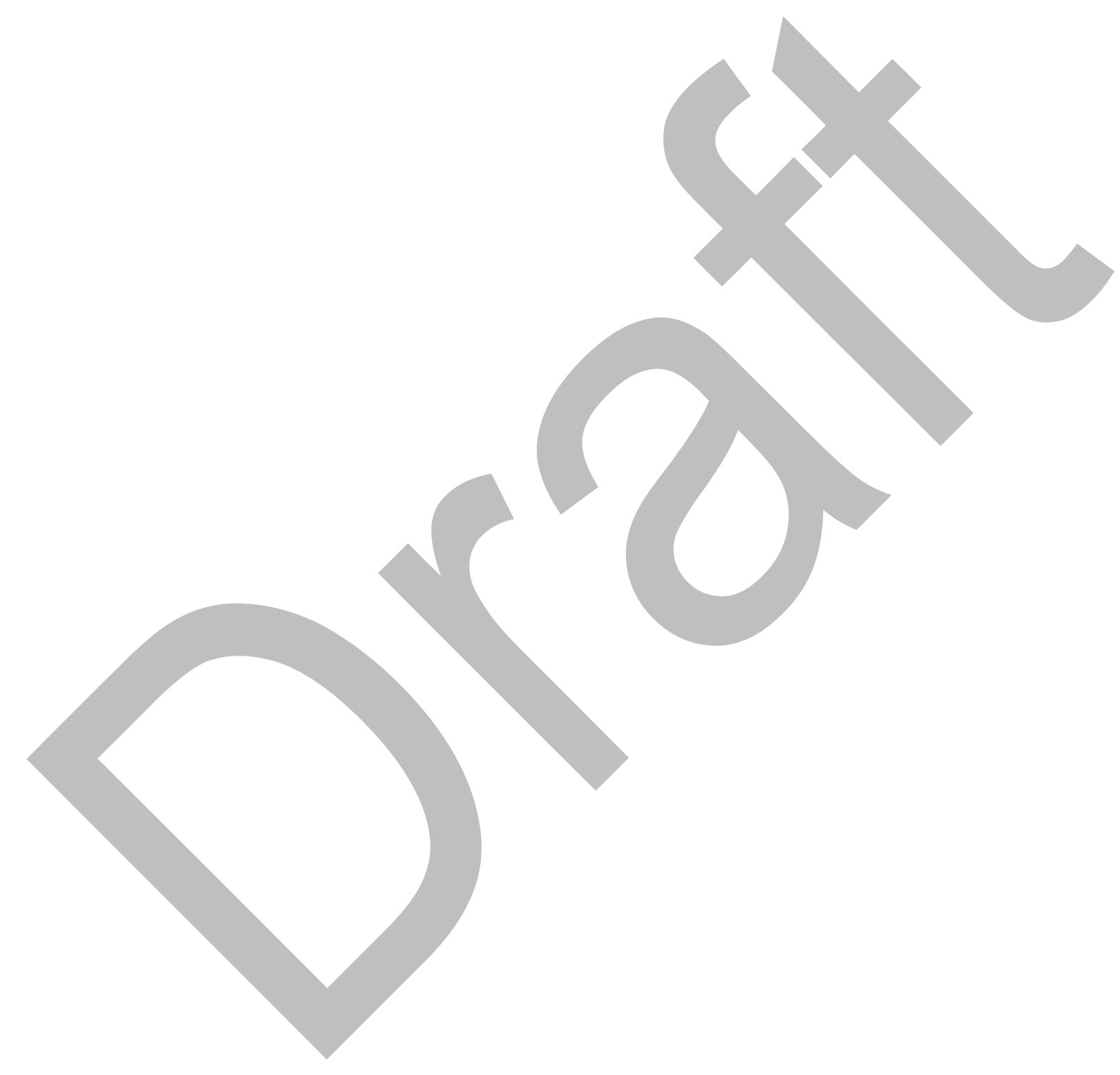

## Analysis of Algorithm

The binary sort is linear in both space and time complexity. The algorithm requires no extra storage space, or memory and performance time is proportional to the number of data elements.

### Space Analysis

The binar sort is an in-place sorting algorithm; hence the same initial array is used for each recursive invocation of the algorithm. The sub-arrays are determined by the passed parameters as boundaries within the array. The only change is for the use of variables for each recursive call of the binar sort.

The array of *N* elements and a constant number of variables *c* for reach recursive invocation is the space complexity, or:

$$O(S) = N + c$$

As there are multiple invocations through recursion, the constant is the sum of each recursive call, or for *i* total calls recursively:

$$O(S) = N + (c_0 + \ldots + c_{i-1})$$

The expression can be simplified using summation notation to the expression:

$$O(S) = N + \sum_{j=0}^{i-1} c_j$$

But the summation of a constant is the product of the constant, or:

$$\sum_{j=0}^{i-1} c_j = i \cdot c_j = c'$$

Effectively the sum of the constants *c* for *i* recursive calls is simply another constant $c'$, or in Big-Oh notation a constant *c*, so the space complexity for the binar sort is:

$$O(S) = N + c = O(N)$$

The space complexity of the binar sort is linear or *O(N)*, proportional to the number of data elements.

### Performance Time Analysis

The analysis of the performance time of sorting algorithms typically considers different cases, from an optimal to a worst case. Each case is simply a different permutation that for the particular sorting algorithm is better or worse in terms of performance. Frequently the permutations of the data, the different cases, are for an ascending sorted permutation, a descending (sorted in reverse) sorted permutation, and a random permutation.

For example, the bubble sort is optimal performance in an almost sorted permutation of the data, and a worse performance for data that is nearly random. The quick sort algorithm is optimal on a random permutation of the data, but is worse in performance for a sorted permutation.

This approach to sorting algorithm analysis is used for the comparison-based sorting algorithms. An unasked question is "Why this approach of various permutations for performance time analysis?" The comparison-based sorting algorithms do a relative comparison for positioning a data element. Each comparison results in a relative positioning of the data element in the overall array of data elements. Thus the initial permutation of the data elements can impact sorting performance.

Why consider this particular point for a sorting algorithm? Simple, the binar sort does not use comparison to place an element in position, but the extracted bit values from a data element. For the constant $c$ total bits in a data element, each bit constitutes $1/c$ in correct positioning a data element. Each positioning is absolute within the overall array, not relative to the other data elements. Effectively, the initial permutation of the data elements is superfluous to the performance time of the binar sort.

In the analysis of the performance time, considering worse and optimal cases is irrelevant as the binar sort is not comparison-based, so is independent of any particular permutation of the data elements.

Analysis of the Algorithm

Analysis of the binar sort algorithm does not use specific cases. The operation of the algorithm impacts the performance. The analysis approach is to consider each discrete step, and consider the total performance time $O(T)$ as the sum of each step. Thus the initial analysis of the performance time of the binar sort is:

$$O(T) = O(T_0) + O(T_1) + O(T_2) + O(T_3)$$

Each discrete step has its own performance time, and for each discrete step the performance time is:

1. $O(T_0)$ - Evaluate for recursive base case.
2. $O(T_1)$ - Initialize starting array bounds.
3. $O(T_2)$ - Partition array into sub-arrays.
4. $O(T_3)$ - Determine recursive call on sub-arrays.

Except for the partition step of the algorithm, all the other steps are constant in the performance time. For the last step involving a logical decision of the recursive call, the complexity of the evaluation is constant for the decision, and the greater performance time for two recursive calls is used as the constant. Substituting a constant for each step, the performance time of the binar sort becomes:

$$O(T) = O(c_0) + O(c_1) + O(T_2) + O(c_3)$$

The expression simplifies the constants to a single constant *c*, and then is:

$$O(T) = O(T_2) + c$$

Partitioning Analysis

The analysis of partitioning is not constant, as the partitioning step in the algorithm is iterative. However, the iterative step processes through all elements in the array, and for the recursive continuation the sub-arrays, which are divisions of the overall array. Thus partitioning is a constant performance time *c* for the bit extraction and determination of which sub-array to place a data element. But this process is repeated n times, or once for each data element. This means that the performance time for the partitioning step is:

$$O(T_2) = c \cdot N$$

Substituting this into the original performance time expression for the algorithm is:

$$O(T) = c \cdot N + c$$

But for the Big-Oh time complexity, this expression simplifies to:

$$O(T) = N$$

Recursion Analysis

One important consideration in the performance time analysis, is that for the expression as the sum of the performance time for each discrete step, is that it is for one operation of the algorithm on a single bit position. This means the performance time expression is:

$$O(T) = 1 \cdot N$$

For each recursive pass, each bit of each data element is accessed once, and there are a constant number *c* of bits (and for other data types with variants, such as a string of characters, there is an overall average length for a data element that is independent of the number of data elements). This means that the recursion will occur for each bit position, so that the performance time is:

$$O(T) = c \cdot N$$

Again, this performance time is linear or *O(N)*.

## Summary of Analysis

Regardless of the permutation of the data elements, the binar sort in the worst case will access a data element once for each bit. Hence for $N$ data elements with a constant or average number of bits $c$, the performance time is $O(N)$. The binar sort algorithm operates independently of the permutation of the data elements, so avoids a worst or optimal case. The binar sort remains linear in both the optimal and the worst case, which precludes any worst or optimal case; the binar sort algorithm remains consistently linear in performance time.

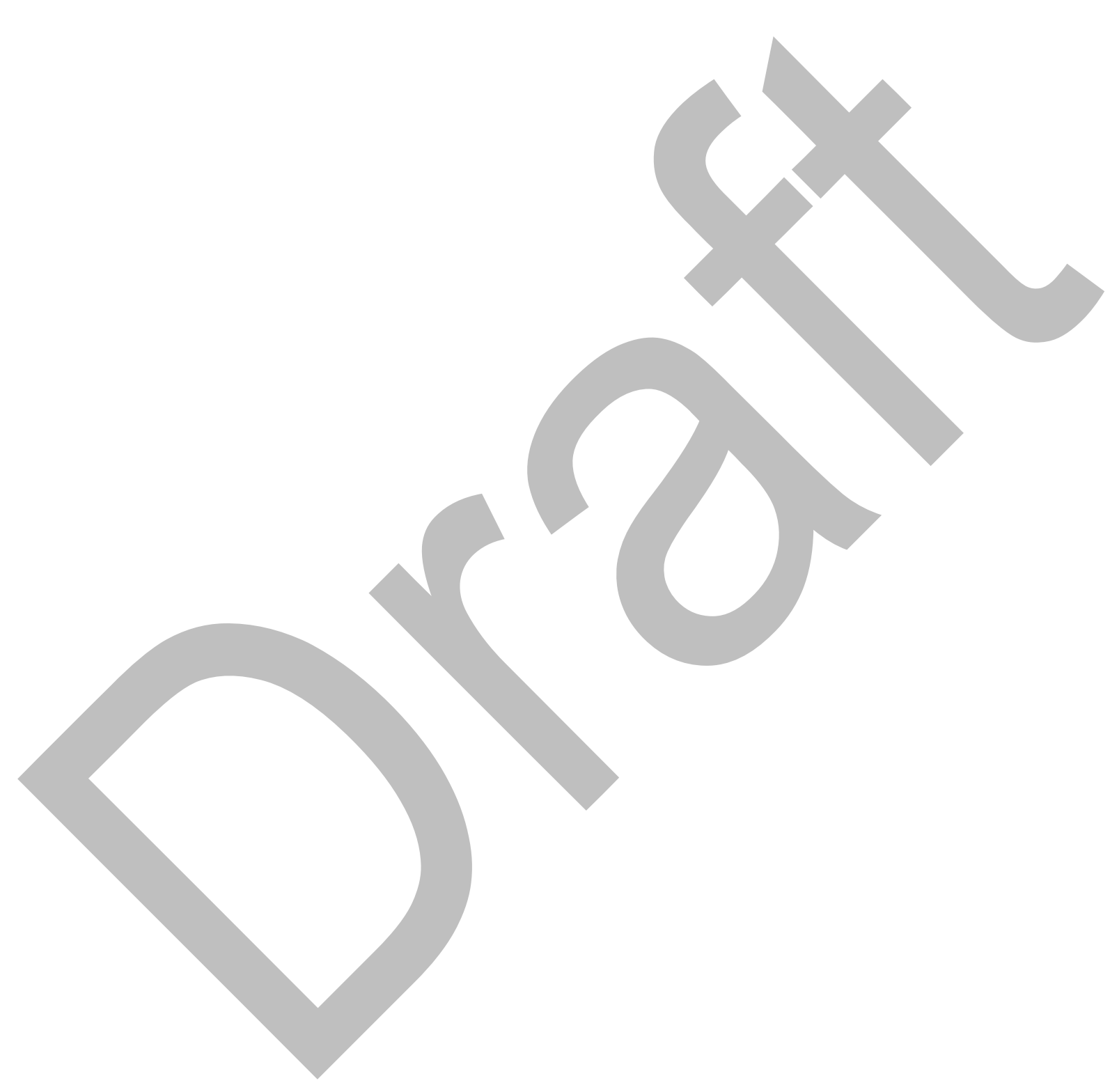

Performance of Algorithm

The theory of the operation and performance time and space complexity of the binar sort are tested by an actual test program written in Java and compiled into Java byte code for the Java Virtual Machine, and using Microsoft J# to execute on the .NET platform.

The test program generates an array of 32-bit integers for a bit size of 32-bits. For test set creation, the Mersenne Twister random number generator is used for initialization of different data sets sizes in random permutations. The test program generates a test size of data from a starting and ending range, with a step in the data size for a performance test.

Performance test parameters are, for example, such as from 10,000 elements to 1,000,000 elements in increments of 1,000 elements. For better accuracy of the tests, a granularity factor is used to run the test and compute the average time, such as 10 or 100 iterations. The goal is to characterize performance with different parameters of the tests, but not to try and calculate some metric of number of integers to time in milliseconds or microseconds.

The test program was executed on different data sizes with varying granularity. The curve shape remains consistent for different test sets, such as 10,000-elements to 1,000,000-elements by 10,000-element steps with a granularity of 100-iterations. One performance test conclusion from the different set of tests was that for 10-iterations of granularity approximately the same as 100-iterations of granularity.

The comprehensive test program uses parameters 1-million integers to 10-million by step of 1000-elements with granularity of 10-iterations. The chart of the performance was created from this particular test, which has the same curve shape as the other tests.

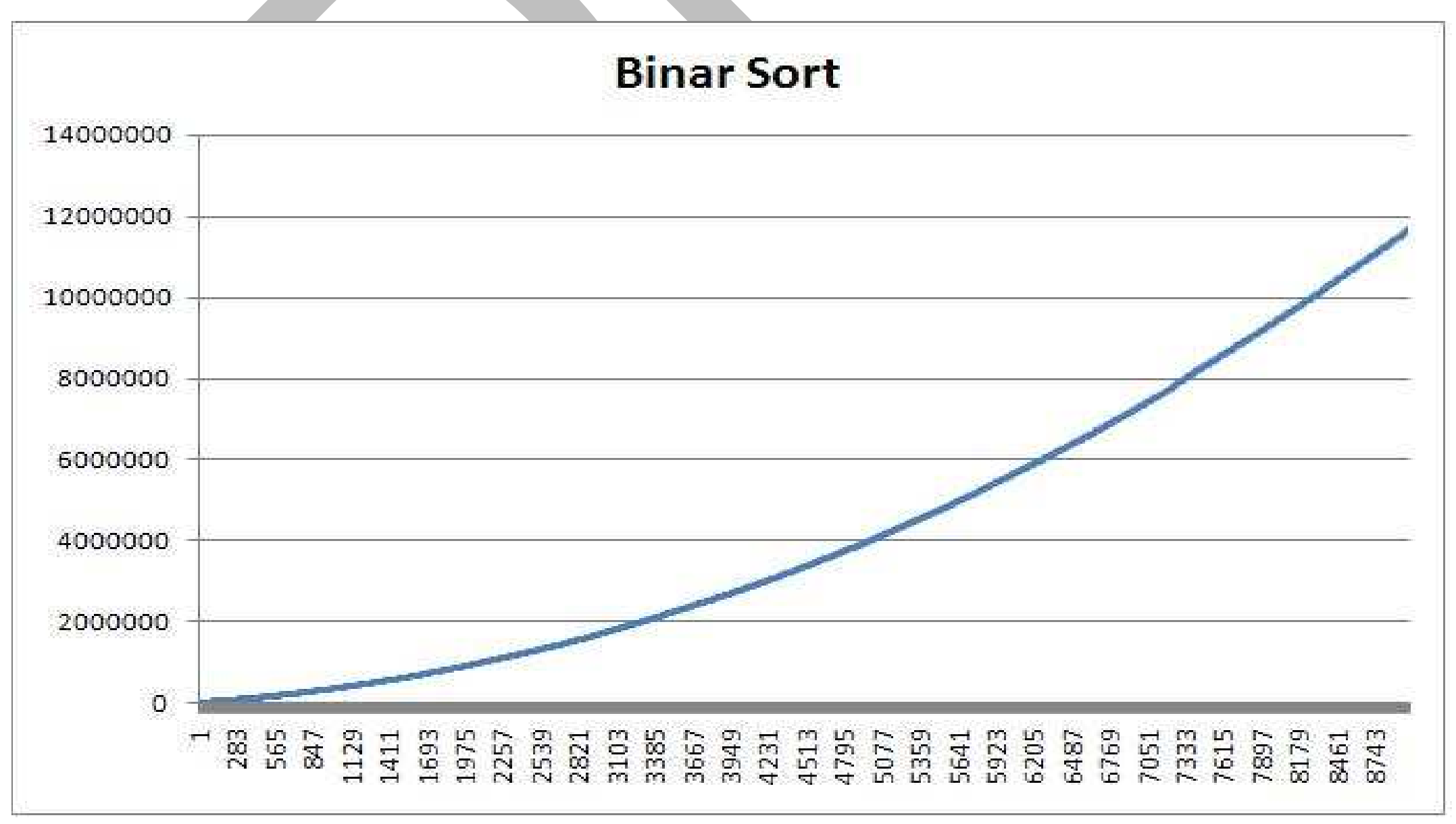

Chart of Test for Binar Sort Performance

The curve shape is not a straight linear plot; in some ways, the plot is sub-linear, a plot of a straight line with a dip or curve beneath the linear line for the performance.

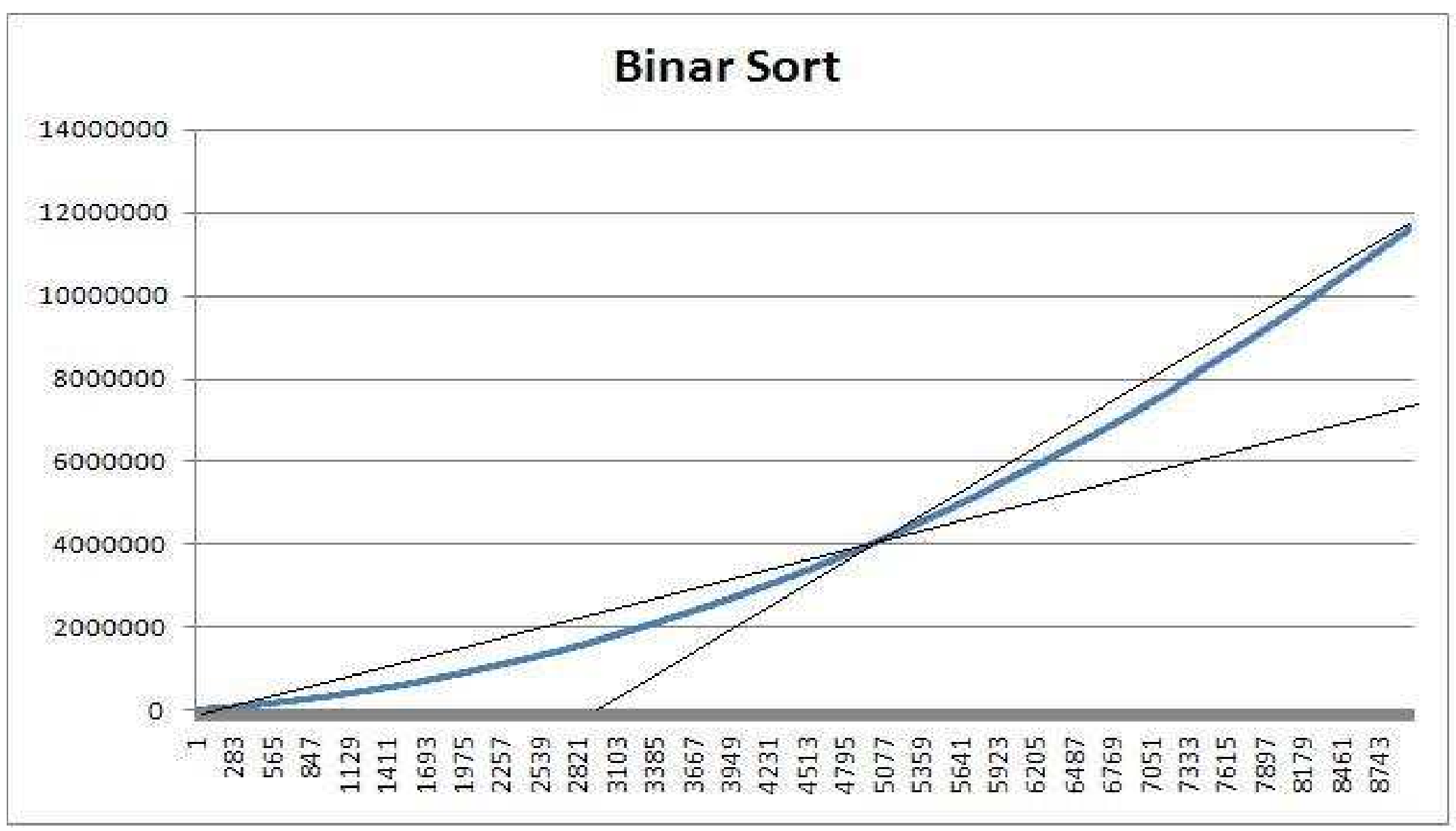

Chart of Test for Binar Sort Performance with Straight Lines Illustrating Linearity

The two major plots of straight lines super-imposed on the plot of the performance time for different data sizes illustrates.

The performance tests show that the binar sort remains linear, although the linear constant of c changes within a range of sizes for the test data. The dip or curve indicates slightly faster performance than linear, hence sub-linear but consistent with the super-imposed straight lines in the chart.

The practical performance time is what is expected, linear time with a constant factor.

## Future Work

The binar sort algorithm is not a finished or completed, there is much work to do with the algorithm. There are two major potential aspects for further work in the future, which are:

1. Variants in the algorithm.
2. Optimize the algorithm.

### Algorithm Variants

The binar sort is recursive, and designed for a serial processor. This leads to the opportunity for:

1. Purely iterative implementation.
2. Parallel version of the algorithm.

The algorithm variant of purely iterative is to replace recursion, if possible, with iteration. The implementation is more an investigation into the potential possibility, so an open research question. An iterative version would remove the overhead in time and space of activation records on the runtime stack of the environment. A comparison of the performance of a recursive to iterative implementation of the binar sort would show possible performance improvement.

The binar sort operates on a serial processor, but it has potential parallelization. A parallel version of the binar sort would let each processor focus on the partition for the processor identity. Thus processor number 9 would focus on the partitioning into 1,0,0,1 and further partition and sort data elements with that bit signature. The interesting feature is that from the initial array of data elements, each parallel operation is completely independent of the other. Theoretically, the linear algorithm of $O(N)$ increases in performance by $O(N/P)$ where there are $P$ processors.

### Optimize the Algorithm

The binar sort is efficient as a linear algorithm, but optimizations are possible in the operation of the algorithm. Two such possible optimizations are:

1. Improve for the partitioning case of pass through.
2. Optimize for determining if elements in sorted order.

One potential optimization is for pass through when partitioning, essentially no data elements are partitioned into sub-arrays. In such a case, the cost of a recursive call to continue partitioning in the same bounds for the next bit position is optimized by iteration. Instead of a recursive call, the partitioning would proceed but for the next bit position. This also simplifies the continuation recursive step, as the recursion would always be for two sub-arrays, not just a single one.

The illustration of the operation of the binar sort demonstrated that the binar sort continues the recursive invocation of the algorithm when the entire array is in a sorted arrangement. This is wasteful of time and the overhead of additional recursion without any further processing to sort needed. A potential optimization is to check within the original array bounds to verify if the ele-ments are in sorted order, and return a logical true or false. The difficulty is to avoid unnecessary checks and the time wasted when the array is not potentially sorted, but to determine when the array is sorted to avoid wasted recursive calls to partition. For example, after several occurrences of pass through while partitioning a sorted order check is performed. If the array is in a sorted ordered arrangement, the recursive calls in other sub-arrays would terminate.

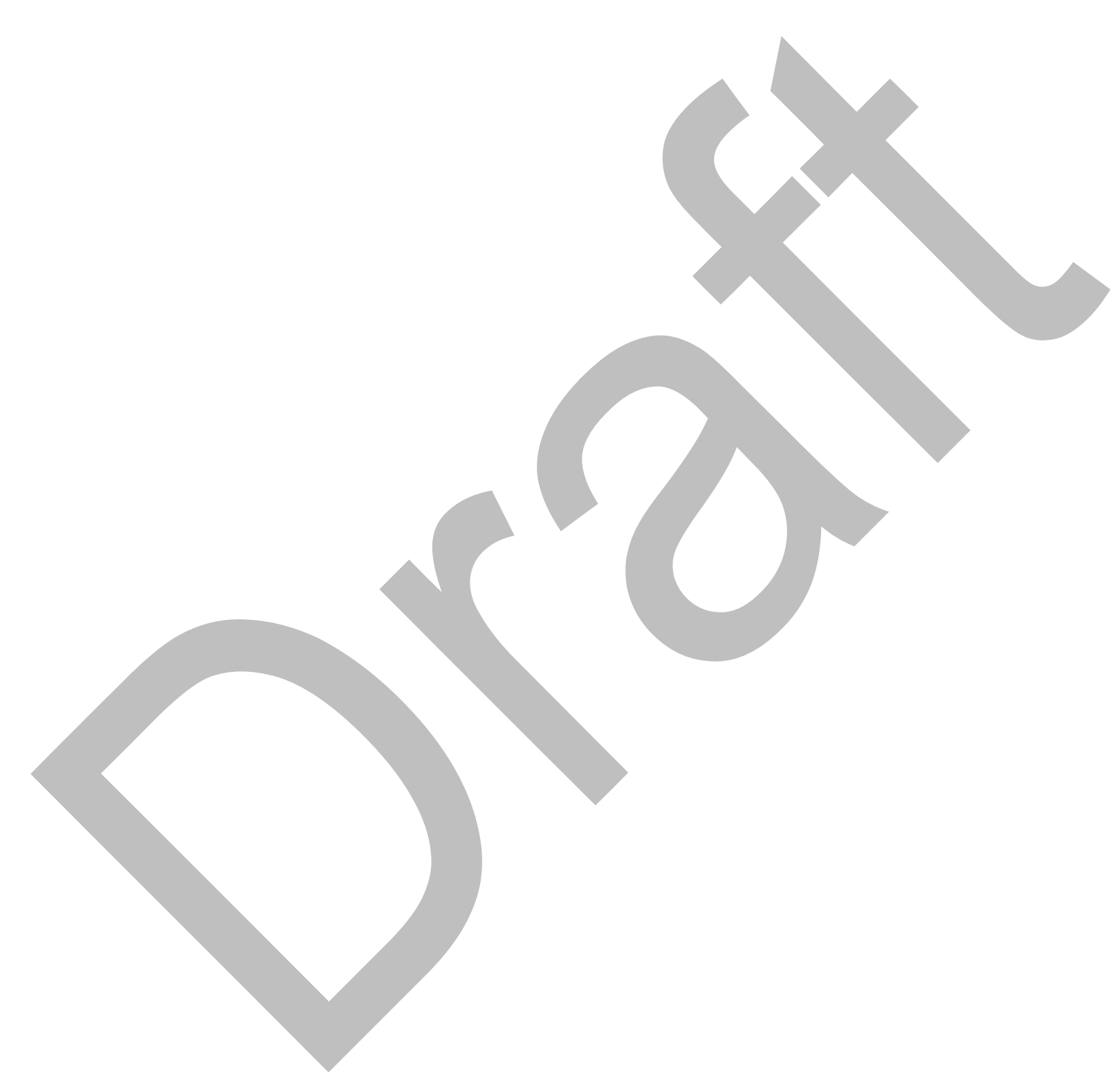

## Conclusion

The binar sort is a linear, unstable, and recursive sorting algorithm. The binar sort algorithm is independent of the initial data permutation. By using the bits of the encoded data elements on a binary computer, the binar sort arranges the data elements into a sorted permutation. Utilizing the encoding makes the binar sort universal, but also a linear algorithm.

Beyond the features of the binar sort algorithm, are some intriguing realizations. One is that generality and performance are not a hard, fast, trade-off that is frequently presumed. The binar sort, by its nature illustrates that such a trade-off is not immutable. The binar sort algorithm illustrates that the linearithmic, comparison based sorting algorithms have an implicit presumption that ordering is the only universal property feasible for generalized sorting. This implicit presumption is flawed, and the other universal property is that for a binary computer (of which all computers at the time of this writing are) the universal property is the encoding of data.

In the analysis of the binar sort algorithm, the permutation of the data is superfluous and irrelevant to the performance of the binar sort. This illustrates that comparison based sorting algorithms are sensitive to the permutation of the data set to sort. By using encoding instead of ordering, the binar sort avoids the sensitivity flaw to the permutation of the data. This flaw in sensitivity to the permutation of the data is an accepted defect in existing comparison based sorting algorithms, but again it is an implicit presumption that this is necessary.

The binar sort algorithm breaks new ground, but it is far from complete or a finished algorithm. Further work exists to optimize the binar sort (including other optimizations that others will find and implement), and alternative implementations. A purely iterative, and an implementation for a parallel computer system are just two areas of future work. The author looks forward with great anticipation at the future work, and possibly other algorithms inspired by the binar sort.

## Appendix Annotated Binar Sort Source Code for Unsigned Integers

```
void BinarSort(int lower, int upper, int pos, int[] array)
{
      //check for termination of recursion to return without partitioning
      if(pos == 33 || upper < lower + 1) return;

      //retain original array boundaries for later use in the code.
      int lo = lower; //array lower bound position
      int hi = upper; //array upper bound position

      //while boundaries do not crossover, continue partition iteration
      //lo increases to hi, hi decrease to lo, crossover with one element
      while (lo < hi + 1){
          //extract the bit from the element, at the lower index position
          //shift element by bit position, and use bit-mask to get bit
          int bit = (array[lo] << pos) & 0x80000000;

          //based upon bit put element into lower or upper sub-array
          if (bit == 0) //bit is 0 move element in lower sub-array
          {
              //element in lower sub-array, extend boundary encompass it
              lo++;
          }
          else //bit is non-0 element in upper sub-array
          {
            //exchange element by swap to lower index position
            int temp  = array[hi];
            array[hi] = array[lo];
            array[lo] = temp;

            //element in upper sub-array, extend boundary to encompass it
            hi--;

          }//end if

      }//end while

    //check if sub-array encompasses entire original array, pass through
    if(lo == upper + 1)
    {
        //pass through, use array bounds at next bit position to continue
        BinarSort(lower, upper, pos + 1, array);
    }
    else
    {
        //for partition, pass sub-arrays at next bit position
        BinarSort(lower, lo - 1, pos + 1, array); //pass lower sub-array
        BinarSort(lo, upper, pos + 1, array);     //pass upper sub-array
    }//end if

}//end BinarSort
```

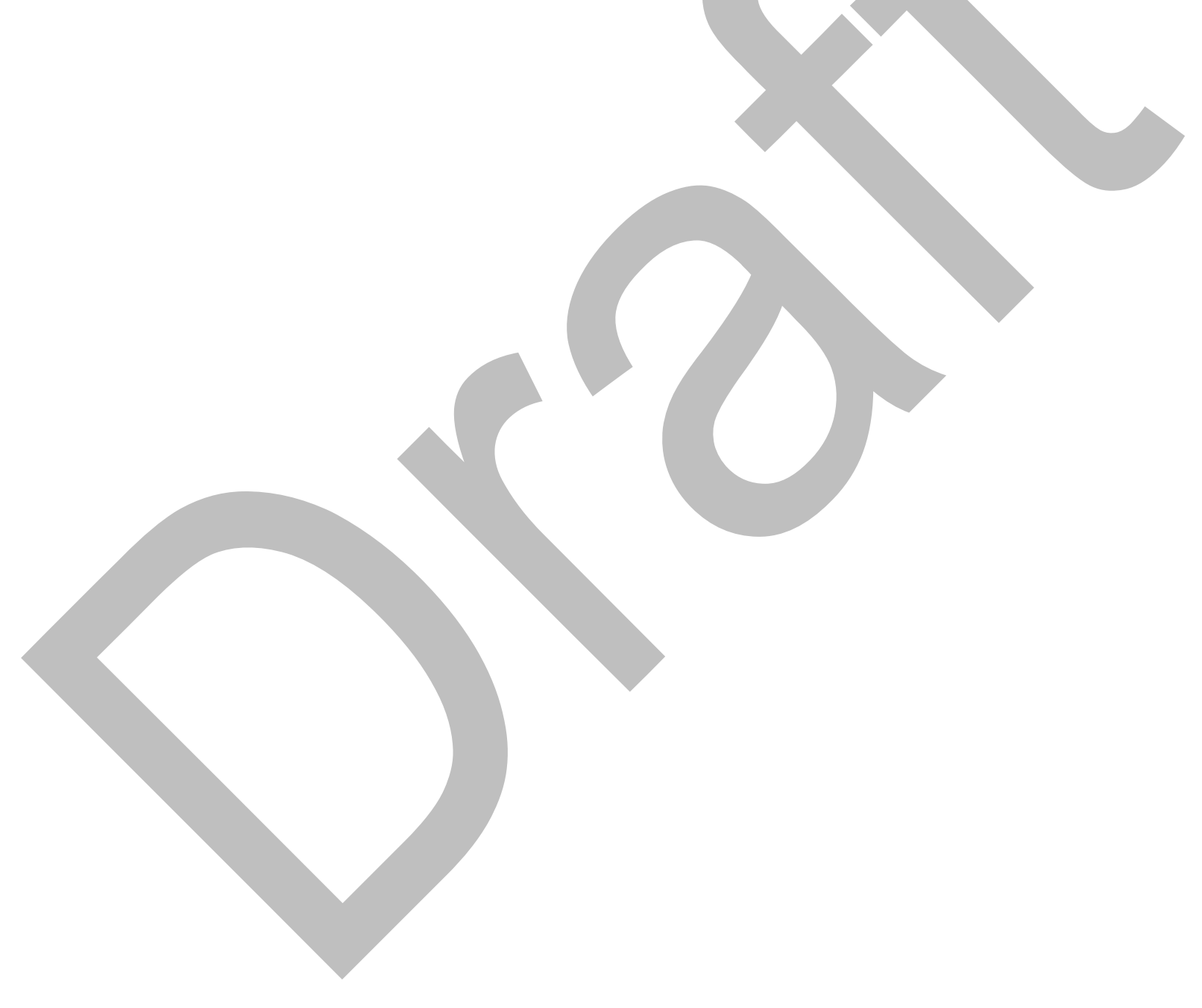